\documentclass[twocolumn]{aastex6.3.1}

\usepackage{amsmath}
\usepackage{float}
\usepackage{booktabs}
\usepackage{hyperref}
\usepackage{verbatim}

\usepackage{tabularx}

\usepackage[]{footmisc}
\usepackage[caption=false]{subfig}

\usepackage{color}

\newcommand{\m}{$m$}

\newcommand{\psrpoppy}{{\sc psrpoppy}}

\newcommand{\psrbh}{PSR--BH}
\newcommand{\rate}{\mathcal{R}}

\newcommand{\rateLigo}{\mathcal{R}_{\rm LIGO}}

\newcommand{\msun}{M$_{\odot}$}

\begin{document}

\title{Insights into the Milky Way pulsar--black hole population using radio and gravitational wave observations}

\author{Nihan Pol}
\affil{Department of Physics and Astronomy, Vanderbilt University, 2301 Vanderbilt Place, Nashville, TN 37235, USA}
\author{Maura McLaughlin}
\author{Duncan R. Lorimer}
\affil{Department of Physics and Astronomy, West Virginia University, Morgantown, WV 26506-6315}
\affil{Center for Gravitational Waves and Cosmology, West Virginia University, Chestnut Ridge Research Building, Morgantown, West Virginia 26505}

\begin{abstract}
    The detection of two NS--BH mergers by LIGO-Virgo provided the first direct confirmation of the existence of this type of system in the Universe.
    These detections  also imply the existence of pulsar--black hole systems.
    In this analysis, we use the non-detection of any \psrbh{} systems in current radio surveys to estimate a 95\% upper limit of $\sim$150 \psrbh{} binary systems that are beaming towards the Earth in the Milky Way. This corresponds to a 95\% upper limit of $\rateLigo = 7.6$~yr$^{-1}$ on the merger detection rate for the LIGO-Virgo network scaled to a range distance of 100~Mpc, which is consistent with the rates derived by LIGO-Virgo. In addition, for the first time, we use the merger detection rates estimate by LIGO-Virgo to predict the number of detectable \psrbh{} systems in the Milky Way. We find there to be $\left< N_{\rm obs, NSBH, e} \right> = 2^{+5}_{-1}$ and $\left< N_{\rm obs, NSBH, p} \right> = 6^{+7}_{-4}$ detectable \psrbh{} systems in the Milky Way corresponding to the event-based and population-based merger detection rates estimated by LIGO-Virgo respectively. We estimate the probability of detecting these \psrbh{} systems with  current radio pulsar surveys, showing that the Arecibo PALFA survey has the highest probability of detecting a \psrbh{} system, while surveys with recently commissioned and planned telescopes are almost guaranteed to detect one of these systems. Finally, we  discuss the hurdles in detecting \psrbh{} systems and how these can be overcome in the future.
    
\end{abstract}

          
\section{Introduction}
    
    Since their discovery more than 50 years ago, pulsars, which are rotating neutron stars (NSs), have been observed in binaries around main sequence stars, white dwarf stars, and other neutron stars \citep{bpsr_lorimer_review}. To date, we have detected more than 300 binary pulsar systems \citep[from the ATNF pulsar catalog,][]{atnf_psrcat}, which include systems such as a double neutron star system with both neutron stars observed as pulsars \citep{Lyne_dpsr, 0737A_disc} and a triple system with a pulsar in an orbit consisting of two other white dwarf companions \citep{triple_system}. These systems have allowed tests of general relativity in the strong-field regime \citep{test_of_GR_dpsr}, constraints on alternative theories of gravity \citep[for e.g.,][]{alt_grav_test_1, triple_system}, insights into magnetospheric physics \citep[for e.g.,][]{dpsr_magneto_1, dpsr_magneto_2}, as well as constraints on different stellar evolution models \citep[for e.g.,][]{bpsr_stellar_evolution, my_nature_paper}. However, despite this diversity in the binary pulsar systems discovered to date, there has been no confirmed detection of a neutron star-- or pulsar--black hole (\psrbh{}) system.
    
    The LIGO-Virgo collaboration recently reported the detection of two NS--BH mergers \citep[][]{ligo_nsbh_detections}. These detections conclusively established the existence of this type of system in the (local) Universe, with the mass of the neutron stars and black holes in the two detections falling between 90\% credible bounds of 1.2~\msun{} and 2.2~\msun{} and 3.6~\msun{} and 10.1~\msun{}, respectively. The merger rate density based on these two detections was estimated to be $\mathcal{R} = 45^{+75}_{-33}$~Gpc$^{-3}$~yr$^{-1}$, which when scaled to a range distance of $D_{r} = 100$~Mpc, results in a merger detection rate of $\mathcal{R}_{\rm event} = 0.19^{+0.31}_{-0.14}$~yr$^{-1}$ \citep[][]{ligo_nsbh_detections}, with the errors representing the 90\% credible region. However, as described in \citet{ligo_nsbh_detections}, if a broader range of priors is assumed for the NS--BH mergers, the corresponding merger rate density is slightly higher at $\mathcal{R} = 130^{+112}_{-69}$~Gpc$^{-3}$~yr$^{-1}$, which scaled to a range distance of 100~Mpc results in a merger detection rate of $\mathcal{R}_{\rm pop} = 0.54^{+0.47}_{-0.28}$~yr$^{-1}$. 
    
    Based on these merger rate estimates, \citet{ligo_nsbh_detections} found that formation mechanisms such as isolated binary evolution \citep[for e.g.,][]{isolated_channel_belczy_1, isolated_channel_belczy_2, isolated_channel_brkgard}, in young stellar clusters \citep[for e.g.,][]{young_cluster_1}, and in disks of active galactic nuclei \citep[][]{agn_formation_channel} all produce merger rate estimates that are consistent with those derived using the observed NS--BH mergers. In the consideration of NS--BH or \psrbh{} systems in the Milky Way, the formation channel involving active galactic nuclei disks is not applicable, while the rate predictions from young stellar clusters, ranging between 0.1--100~Gpc$^{-3}$~yr$^{-1}$ are in slight tension with the larger, population based rate estimate from \citet{ligo_nsbh_detections}. This leaves the isolated binary formation channel as the most viable formation mechanism for NS--BH or \psrbh{} systems.
    Additionally, dynamical formation scenarios \citep[][]{dynamical_formation_pz} produce merger rates that are inconsistent with the observed merger rate, unless special accommodations are made in the stellar evolution processes \citep[for e.g.,][]{globular_cluster_origin_1, globular_cluster_origin_2}. 
    
    Given these different formation channels for NS--BH and \psrbh{} systems, many radio surveys have been designed and executed over multiple decades to target each of these formation channels.
    Finding \psrbh{} systems has been a major objective for all of the large radio pulsar surveys that have been carried out to date \citep[for e.g., ][]{PALFA, aodrift_1, PMSURV, gbncc, htru_low_mid}, whose primary focus is searching along the plane of the Galaxy for systems formed through the isolated binary channel. There have also been a number of targeted radio pulsar surveys of globular and open clusters in the Milky Way \citep{urquhart_globc_survey, meerkat_globc_survey, FAST_globc_survey}, as well as the Galactic center \citep{kaustubh_GC_survey, hyman_GC_survey, effelsberg_GC_survey}, which aim to detect systems formed through both the isolated binary and dynamical formation channels. However, none of these surveys have reported a detection of a \psrbh{} system.
    
    While the two NS--BH merger detections made by LIGO-Virgo provided invaluable scientific results about stellar evolution, component masses and even tests of general relativity \citep[GR,][]{ligo_nsbh_detections}, detecting a \psrbh{} system in the Milky Way stands to be just as valuable. A pulsar, as a regular rotator, can be used as a clock in orbit around the black hole in the \psrbh{} system and timing observations of the pulsar will allow direct tests of the space-time surrounding the black hole, as well as probing the properties of the black hole itself. For example, \citet{wex_psrbh_science} showed that detecting a \psrbh{} system would allow us to measure the mass of the black hole to an accuracy better than $\sim$5\%, while measurement of orbital precession can allow us to estimate the spin of the black hole. \citet{liu_psrbh_strongfield_test} and \citet{liu_psrbh_ska_tests} showed that \psrbh{} systems can be excellent sources to test the behaviour of GR in the strong-field regime, complementary to the tests that are possible with gravitational wave merger detections made by LIGO-Virgo. \citet{liu_psrbh_ska_tests} also showed that \psrbh{} systems provide excellent tests of alternative theories of gravity, even if those theories predict black holes with identical properties to those predicted by GR. 
    
    Given the wealth of science offered by \psrbh{} systems, it is worth aggregating the information available so far through radio surveys and gravitational wave observations to infer the size of the population and detection prospects for \psrbh{} systems in the Galaxy. In Sec.~\ref{sec:radio_constraint}, we describe the process, and estimate and compare the NS--BH merger detection rate using the non-detection of \psrbh{} systems in current radio surveys. In Sec.~\ref{sec:gw_prediction}, we develop and implement the methodology for using the merger detection rate estimated by gravitational wave observations of NS--BH mergers to estimate the Milky Way population of \psrbh{} systems. In Sec.~\ref{sec:future_prospects} discuss the prospects for detecting these systems as well as hurdles involved in trying to find these systems. In Sec.~\ref{sec:conclusion} we offer our conclusions.

\section{Radio survey constraints on \psrbh{} population} \label{sec:radio_constraint}
    
    \subsection{\psrbh{} population and survey parameters} \label{subsec:pop_prop}

        We use the \psrpoppy{} software package \citep[][]{psrpoppy} to simulate the pulsar population and radio surveys in this analysis. Since we are modelling \psrbh{} binaries, we model the corresponding degradation in the S/N ratio due to the orbital motion of the pulsar by using the neural network implementation of the orbital degradation factor \citep{Bagchi_odf} described in \citet{my_ucb_pop}.
        
        The mass of the pulsars is assumed to follow the distribution given in \citet{ns_mass_dist} for canonical (long-period) pulsars and the mass of the companion is assumed to be uniform between 3~\msun\ and 100~\msun\ \citep{compas_psrbh_debatri}. 
        We assume that the inclination is uniform in sin($i$), where the inclination angle, $i$, is allowed to vary between $0^{\circ}$ and $90^{\circ}$, angle of periastron passage is uniform between $0^{\circ}$ and $360^{\circ}$, and eccentricity uniform between 0 and 0.9. We allow the orbital period to vary between $10^{-3}$~days and 10~days, where the upper limit is chosen such that all systems merge, on average, within a Hubble time.
        For the calculation of the orbital degradation factor, we use only the second harmonic, assuming that the pulsar is an orthogonal rotator and hence, most of its power is contained in the second harmonic \citep[][]{my_merger_rate, my_ucb_pop}.
        
        Since most of the formation channels for \psrbh{} binaries predict that the NS is born after the BH in the system \citep[for e.g.,][]{kruckow_bh_form_first}, the pulsar is unlikely to undergo any period of accretion from the companion, and will be a canonical pulsar. Thus, we assume the pulsar spin periods to have the same distribution as that for the canonical pulsar distribution, i.e. a log-normal distribution with $\left< \log_{10}P \right> = 2.7$ and $\sigma_{\log_{10}P} = 0.34$ \citep[][]{lorimer_rad_dist}. We also assume a log-normal luminosity distribution with mean $\left< \log_{10}L \right> = -1.1$ ($L = 0.07 \, {\rm mJy \, kpc^2}$) and standard deviation $\sigma_{\log_{10}L} = 0.9$ \citep{fk06}. We also consider surveys at different radio frequencies, and thus assume a normal spectral index distribution with mean $\alpha = 1.4$ and standard deviation $\beta = 1$ \citep{bates_si_dist}. We assume the radial distribution of pulsars in the Galaxy to follow that from \citet{lorimer_rad_dist}, and the $z$-height distribution as described by a two-sided exponential with a scale height of $z_0 = 330$~pc.
        
        For canonical pulsars, the beaming fraction, $F$, which represents the percentage of the sky covered by a pulsar's beam, can be related to the spin period \citep{tm_98},
        \begin{equation}
            \displaystyle F = 9 \left( \log_{10} \frac{P}{10} \right)^2 + 3,
            \label{eq:beam_fraction}
        \end{equation}
        where $P$ is the spin period of the pulsar in seconds. The beaming correction factor, $f_{\rm b} = 100 / F$, corrects for the number of systems whose emission beam does not cross the line-of-sight to Earth. 
        
        Finally, we need to know the distribution of the effective lifetime, $\tau_{\rm life}$, of the \psrbh{} system. The effective lifetime is defined the amount of time for which the \psrbh{} binary is detectable through the pulsar in the system. Thus, for the canonical pulsars in this analysis, $\tau_{\rm life} = \tau_{\rm c} + {\rm min}(\tau_{\rm death}, \tau_{\rm merger})$, where $\tau_{c}$ is the characteristic age of the pulsar, $\tau_{\rm death}$ is the time after which the pulsar crosses into the so-called ``death valley'' \citep{Chen_ruderman_death_valley, zhang_death_line} and ceases to be radio-visible, and $\tau_{\rm merger}$ is the time until the \psrbh{} system merges through the emission of gravitational waves.
        Since we use the ``snapshot'' modelling method of \psrpoppy{} in our analysis \citep{psrpoppy}, we do not have knowledge of the spin period derivative of our simulated pulsars, preventing us from calculating the characteristic ages of the pulsars in our simulation. In addition, there is large uncertainty associated with when pulsars cross into the ``death valley'' \citep{zhang_8p5_s_pulsar, 23p5_s_pulsar}.
        Thus, for this analysis, we choose to adopt a log-uniform probability distribution for the effective lifetime of the \psrbh{} systems in our simulations, with the binaries allowed to have lifetimes between 1~Myr and 1~Gyr.

        \begin{table*}
            \centering
            \caption{This table lists the telescope and survey parameters for the large pulsar surveys that are considered in this work.}
            \begin{tabular}{ccccccc}
                \toprule
                Survey & Gain, $G$ & Center Frequency, $f_{\rm c}$ & Bandwidth, $B$ & System temperature, $T_{\rm sys}$ & Integration time, $t_{\rm int}$ \\
                 & (K/Jy) & (MHz) & (MHz) & (K) & (s) \\
                \toprule
                & & & \textit{Current Surveys} & & \\
                \midrule
                PALFA\footnote{Pulsar Arecibo L-band Feed Array, \citet{PALFA}} & 8.5 & 1374 & 300 & 25 & 268 \\
                PMSURV\footnote{Parkes Multibeam SURvey, \citet{PMSURV}} & 0.6 & 1374 & 288 & 25 & 2100 \\
                AO327\footnote{Arecibo DRIFT scan survey, \citet{aodrift_1}} & 10 & 327 & 25 & 100 & 50 \\
                GBNCC\footnote{Green Bank North Celestial Cap Survey, \citet{gbncc}} & 2 & 350 & 100 & 46 & 120 \\
                HTRU--low\footnote{High Time Resolution Universe low-latitude survey \citet{htru_low_mid}} & 0.6 & 1352 & 340 & 25 & 340 \\
                HTRU--mid\footnote{High Time Resolution Universe mid-latitude survey \citet{htru_low_mid}} & 0.6 & 1352 & 340 & 25 & 540 \\
                \midrule
                & & & \textit{Future Surveys} & & \\
                \midrule
                FAST GPPS\footnote{Five hundred meter Aperture Spherical Telescope \citep{FAST} Galactic Plane Pulsar Survey, \citet{FAST_gpps}} & 16 & 1250 & 450 & 25 & 300 \\
                DSA-2000 Pulsar survey\footnote{Deep Synoptic Array, \citet{dsa2000}} & 2 & 1300 & 1300 & 25 & 600 \\
                MeerKAT Pulsar Survey\footnote{\citet{meerkat_psr}} & 2.8 & 1284 & 776 & 18 & 600 \\
                \bottomrule
            \end{tabular}
            \label{tab:surveys}
        \end{table*}
        
        The surveys that are used in our simulations are listed under ``Current surveys'' in Table~\ref{tab:surveys}. These surveys represent the largest radio pulsar surveys that have been carried out so far. We make the same assumption as in \citet{my_ucb_pop}, where we assume that these surveys will complete their design sky coverage. This, unfortunately, will not be possible for the surveys that were being carried out at the Arecibo radio telescope, namely the PALFA and AO327 surveys, due to its recent collapse, but we nevertheless assume that that these surveys have covered $>90\%$ of their designed sky coverage \citep{PALFA_completeness, AODRIFT_completeness}.
        
    \subsection{Statistical framework} \label{subsec:stats_framework}
    
        In this approach, we use the the statistical framework developed by \citet{kkl} and used most recently in \citet{my_ucb_pop} and \citet{dns_merger_rate_kramer}, and we refer the reader to them for details about the framework.

\begin{figure}
            \centering
            \includegraphics[width = \columnwidth]{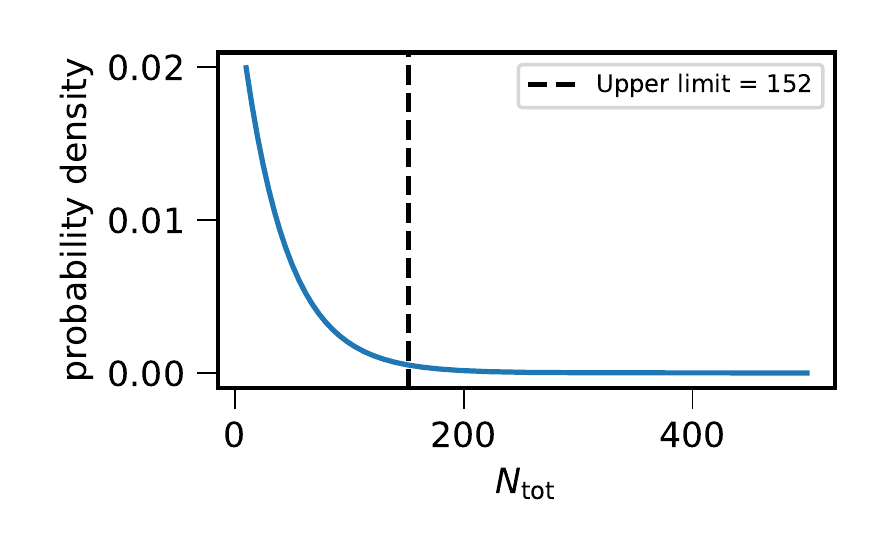}
            \caption{The 95\% credible upper limit on the number of \psrbh{} systems that are beaming towards the Earth given that none of the current radio surveys have detected one of these systems so far.}
            \label{fig:pop_size}
        \end{figure}

        Similar to \citet{my_ucb_pop}, since no \psrbh{} systems have been detected, we set $N_{\rm obs} = 0$ in our analysis. The corresponding 95\% confidence upper limit on the number of \psrbh{} systems that are beaming towards Earth, $N_{\rm tot}$, is shown in Fig.~\ref{fig:pop_size}.
        
        \begin{figure}
            \centering
            \includegraphics[width = \columnwidth]{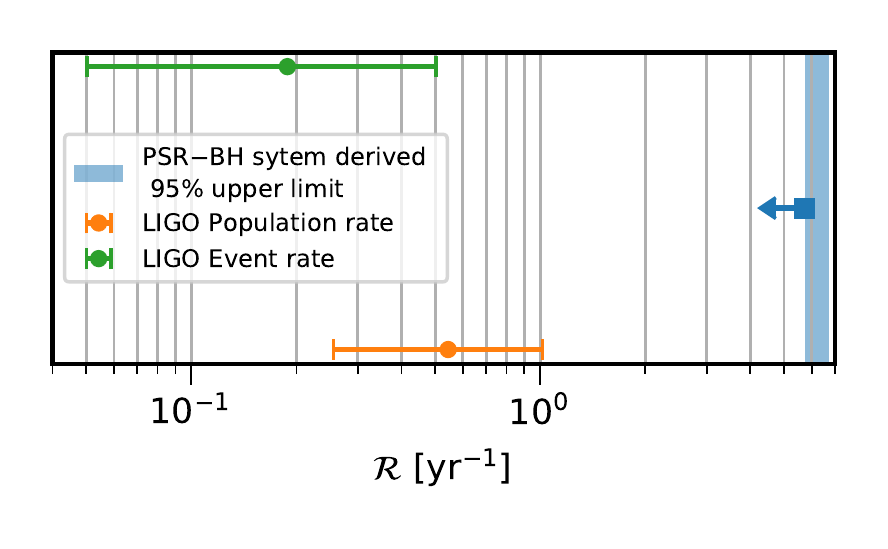}
            \caption{The 95\% credible upper limit on the merger detection rate based on the non-detection of a \psrbh{} system in current radio surveys is shown along with the merger detection rates derived using LIGO-Virgo detections of two merging NS--BH systems.}
            \label{fig:rates}
        \end{figure}
        
        We convert this to the corresponding prediction of the LIGO merger detection rate \citep{kkl, kopparapu_extrapolate} and derive a 95\% confidence upper limit on the merger detection rate for LIGO of $\rateLigo = 7.6$~yr$^{-1}$. This merger detection rate is shown in Fig.~\ref{fig:rates}, along with the merger detection rates derived in \citet{ligo_nsbh_detections}. 
        This merger detection rate is consistent with that derived by \citet{ligo_nsbh_detections}, which suggests that our assumption of no \psrbh{} system detections in the Milky Way is compatible with the detection of two NS--BH mergers by LIGO-Virgo.
        
\section{Gravitational Wave prediction of the Galactic \psrbh{} population} \label{sec:gw_prediction}

    \subsection{Maximum likelihood Framework} \label{subsec:ml_framework}
        
        The statistical framework of \citet{kkl} assumes knowledge of the number of observed \psrbh{} systems, $N_{\rm obs}$, in order to derive upper limits on the size of the \psrbh{} population and the corresponding merger rate. However, we can leave $N_{\rm obs}$ as a free parameter and use the merger rate estimated by \citet{ligo_nsbh_detections} to place constraints on the number of observable \psrbh{} systems in the Milky Way. Once we know the number of observable systems, we can follow the same process as in Sec.~\ref{sec:radio_constraint} to derive the size of the Milky Way \psrbh{} population.
        
        Following the same derivation as in \citet{kkl} but with $N_{\rm obs}$ as a free parameter and ignoring any normalization factors, the probability distribution for the size of the \psrbh{} population can be written as,
        \begin{equation}
            \displaystyle P(N_{\rm tot}) =
            \frac{
            \alpha^{N_{\rm obs} + 1} N_{\rm tot}^{N_{\rm obs}} e^{-\alpha N_{\rm tot}} }{ N_{\rm obs}!},
            \label{eq:general_pop_size}
        \end{equation}
        while the probability distribution for the merger rate,
        \begin{equation}
            \displaystyle P(\rate) = \left[ \frac{\alpha \tau_{\rm life}}{f_{\rm b}} \right]^{N_{\rm obs} + 1}
            \frac{
            \rate^{N_{\rm obs}} e^{-(\alpha \tau_{\rm life} / f_{\rm b}) \rate}}{ N_{\rm obs}!}.
            \label{eq:general_merger_rate}
        \end{equation}
        In addition to $N_{\rm obs}$, the lifetime, $\tau_{\rm life}$, and the beaming correction factor, $f_{\rm b}$, of the \psrbh{} systems are also free parameters in Eq.~\ref{eq:general_merger_rate}. However, these two parameters are fully covariant with each other due to the structure of Eq.~\ref{eq:general_merger_rate}. As a result, we choose to model the ratio of these parameters, $\beta = f_{\rm b} / \tau_{\rm life}$, rather than the individual parameters themselves. Note that $\alpha$ can be considered as a constant in Eq.~\ref{eq:general_merger_rate} since it depends primarily on the surveys used to search for the \psrbh{} systems.
        
        We use a non-linear maximum likelihood approach to find the values of $N_{\rm obs}$ and $\beta$. The residual function is defined as the difference between the probability density obtained from Eq.~\ref{eq:general_merger_rate} (appropriately normalized) and the probability density function derived using LIGO's observation of NS--BH mergers \citep{ligo_nsbh_detections}. 
        The non-linear optimization is performed using \textsc{lmfit} \citep{lmfit}, which is a Python package providing a high-level interface for optimization methods. \textsc{lmfit} converts the residual function defined above into the corresponding $\chi^2$ distribution by calculating the sum of the squares of the residual function values. We use the Levenberg-Marquardt \citep{levenberg_marquardt} method for minimizing this $\chi^2$ distribution. 
        Since \textsc{lmfit} allows setting bounds on the parameters in Eq.~\ref{eq:general_merger_rate}, we constrain $N_{\rm obs}$ to the range [0, 100], and constrain $\beta$ to the range [$1 \times 10^{-3}$, $10$]. The bounds on $\beta$ were chosen such that they correspond to bounds of [1, 10] and [1~Myr, 1~Gyr] on $f_{\rm b}$ and $\tau_{\rm life}$ respectively. Once the maximum likelihood solution is found, we explore the parameter space around the maximum likelihood solution using the \textsc{emcee} \citep{emcee} sampler provided in \textsc{lmfit}, thereby calculating the 95\% credible regions around the maximum likelihood solution.
        
    \subsection{Testing the framework} \label{subsec:test_ml_framework}
        
        To test this method, we first deploy it on the merging double neutron star (DNS) population in the Galaxy. There have been several recent studies that used the observed population of DNS systems in the Galaxy to predict the DNS merger detection rate for LIGO \citep[e.g., ][]{my_merger_rate, dns_merger_rate_kramer}. We use the DNS merger detection rate estimate from \citet{dns_merger_rate_kramer} as our target function in the non-linear optimization method described above. We can then judge the performance of our method by comparing the predicted value of $N_{\rm obs}$ with the real, known value of the number of merging DNS systems that have been observed so far (nine).
        
        To derive $\alpha$ for the DNS population, we use the same process as described in Sec.~\ref{subsec:stats_framework}. The DNS population properties are mostly similar to those described in Sec.~\ref{subsec:pop_prop} with some key differences. The mass of the secondary (and primary) is assumed to follow the distribution of DNS systems given in \citet{ns_mass_dist}. The majority of pulsars in DNS systems have been observed as millisecond pulsars, but due to the dearth of these systems, the underlying period distribution is still unknown. Thus, we assume a uniform distribution of spin periods, ranging from 1~ms to 100~ms. Similarly, we assume a uniform distribution of orbital periods to correspond to the observed DNS systems \citep[see Table~1 in ][]{my_merger_rate}. The other population properties are the same as described in Sec.~\ref{subsec:pop_prop}.
        
        \begin{figure*}
            \centering
           \subfloat[]{\includegraphics[width= \columnwidth]{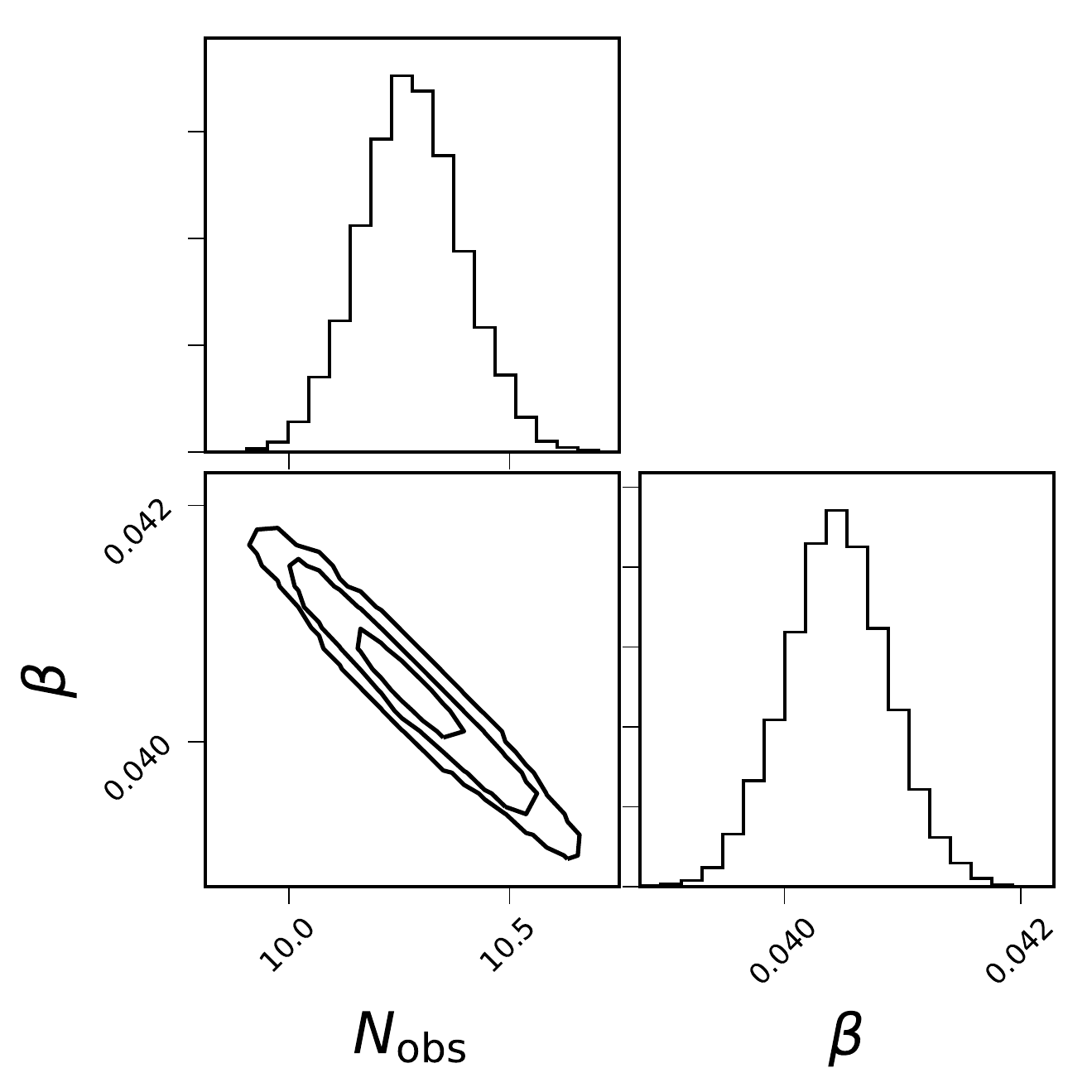}}
            \hfill
            \subfloat[]{\includegraphics[width= \columnwidth]{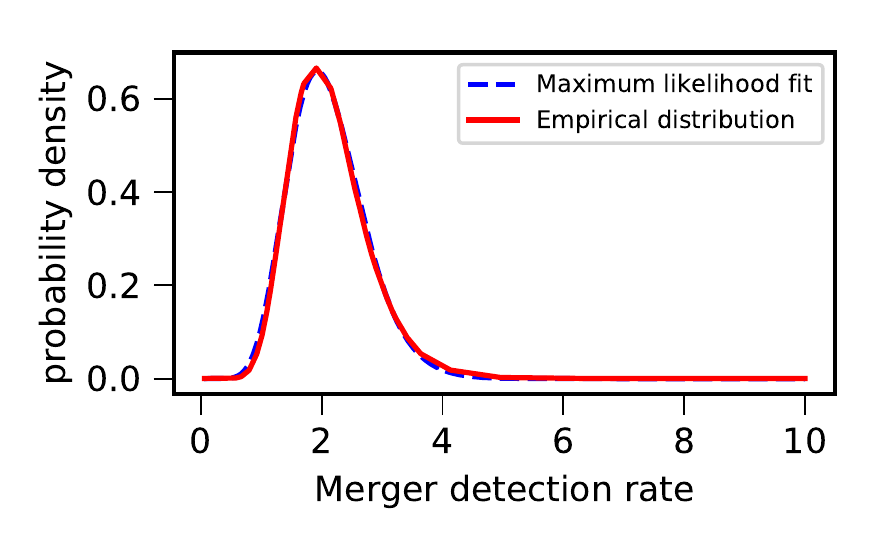}}
            \hfill
            \subfloat[]{\includegraphics[width= \columnwidth]{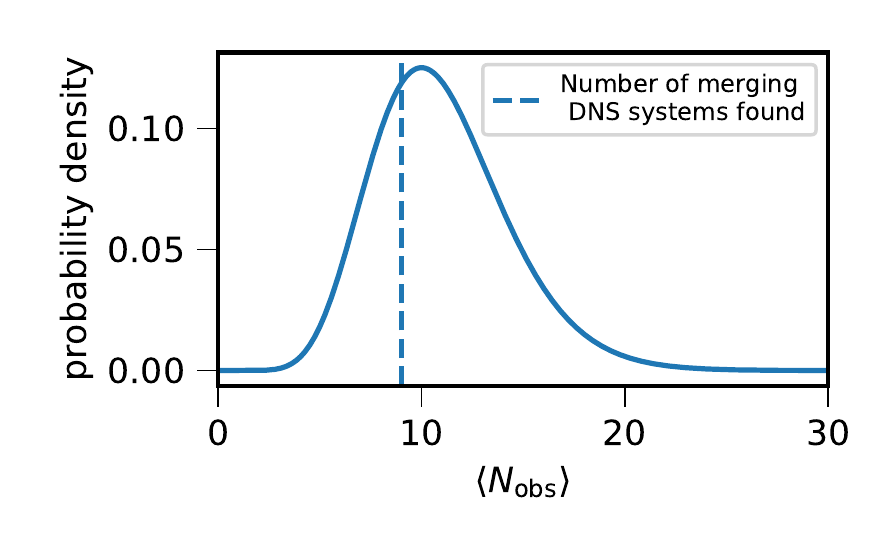}}
            \caption{Results from the maximum likelihood analysis for the population of merging DNS systems in the Galaxy.
            \textit{(a):} The posterior distribution on the parameters in Eq.~\ref{eq:general_merger_rate}. The contours represent the 1-, 2-, and 3-$\sigma$ confidence intervals.
            \textit{(b):} The probability distribution generated using the maximum likelihood parameters shown in panel (a) is compared with the empirical distribution derived in \citet{dns_merger_rate_kramer} showing excellent agreement between the two distributions.
            \textit{(c):} Distribution of the expected number of observable DNS systems from the maximum likelihood analysis described in Sec.~\ref{subsec:ml_framework}. The vertical dashed line shows the number of DNS systems that will merge within a Hubble time that have been discovered in radio surveys so far (nine).}
            \label{fig:dns_max_lkl}
        \end{figure*}

        With this simulation setup, we calculate $\alpha_{\rm DNS}$, which is then used with the framework described in Sec.~\ref{subsec:ml_framework}. To obtain the target merger detection rate distribution, we use the Galactic merger rate of $\mathcal{R} = 32^{+19}_{-9}$~Myr$^{-1}$ \citep[][]{dns_merger_rate_kramer} and convert it to the LIGO merger detection rate, scaled to a range distance of 100~Mpc \citep[][]{kopparapu_extrapolate}. This merger detection rate, along with the best fit curve obtained from our optimization method is shown in Fig.~\ref{fig:dns_max_lkl}. The maximum likelihood values thus obtained are $N_{\rm obs, DNS} = 10.3 \pm 0.23$ and $\beta = 0.0404 \pm 0.0008$. 
        
        The corner plot showing the posterior distribution of these parameters and the distribution of $\left<N_{\rm obs} \right>$ \citep[Eq.~7 in][]{kkl} is also shown in Fig.~\ref{fig:dns_max_lkl}. 
        The recovered distribution of $\left<N_{\rm obs} \right> = 10^{+8}_{-4}$ is consistent with the nine observed DNS systems in the Galaxy. 
        Additionally, even though we cannot set independent constraints on $f_b$ or $\tau_{\rm life}$, we can use the knowledge of $f_b$ for the observed DNS systems from \citet{dns_merger_rate_kramer} to estimate the average lifetime of the Galactic DNS population using the maximum likelihood value of $\beta$. As shown in \citet{my_merger_rate} and \citet{dns_merger_rate_kramer}, there is a small group of DNS systems in the Galaxy that dominate the calculation of the merger detection rate. Using the average beaming correction factor of just these systems \citep[Table 2 in][]{dns_merger_rate_kramer}, $f_{b, \rm avg} = 4.5$, we get the corresponding average lifetime of $\tau_{\rm life, avg} = 111.4 \pm 0.3$~Myr. This estimate is close to the average lifetime of 129~Myr for the group of DNS systems that contribute most to the merger detection rate.
        
        This test shows that the optimization method presented in Sec.~\ref{subsec:ml_framework} can accurately predict the number of observable binary pulsar systems, as well as provide reasonable estimates of the average lifetime of the population under consideration given some knowledge of the beaming correction factor.
        
    \subsection{Estimating the number of observable \psrbh{} systems in the Galaxy} \label{subsec:n_psrbh_obs}
        
        Given the verification of the maximum likelihood framework in Sec.~\ref{subsec:test_ml_framework}, we can apply this method to estimate the number of \psrbh{} systems in the Galaxy. We produce estimates using both the event- and population-based merger detection rates calculated by \citet{ligo_nsbh_detections}. We scale these merger detection rates reported in \citet{ligo_nsbh_detections} to a range distance of 100~Mpc, which are then used as the target function in the maximum likelihood framework. We use the same $\alpha_{\rm NSBH}$ that was derived with the assumptions in Sec.~\ref{sec:radio_constraint}.
        
        \begin{figure}[h]
            \centering
           \subfloat[]{\includegraphics[width= 0.5\textwidth]{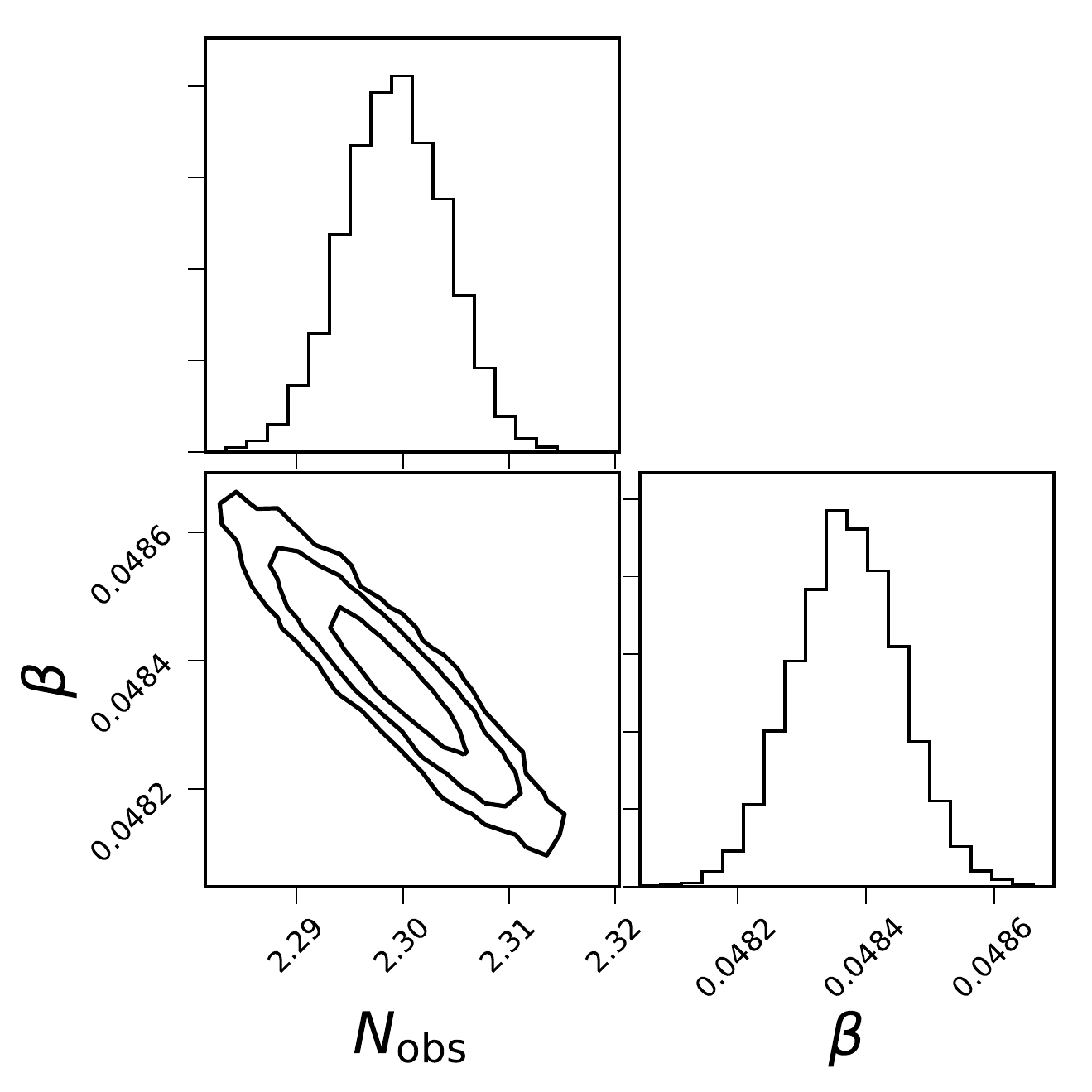}}
            \hfill
            \subfloat[]{\includegraphics[width= \columnwidth]{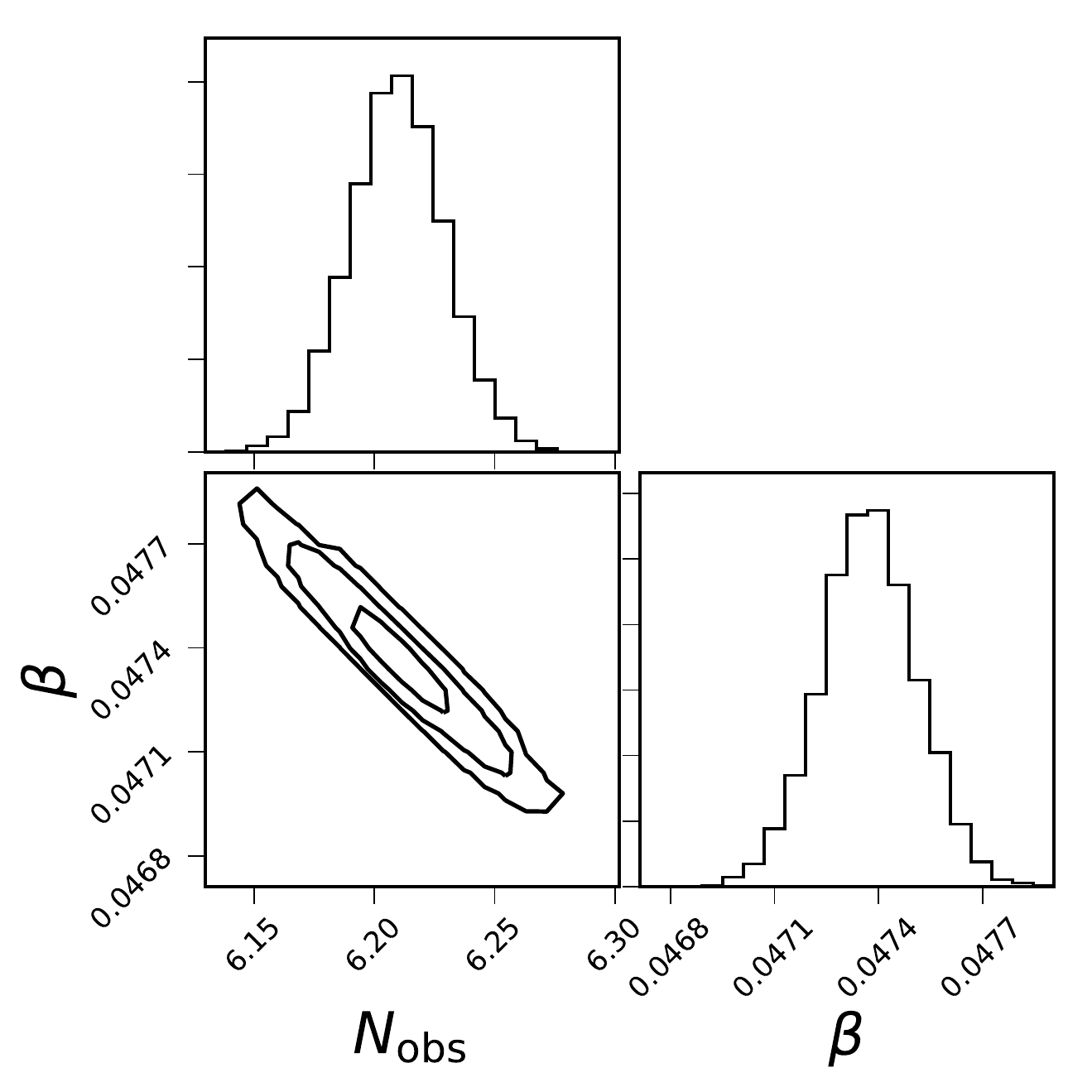}}
            \caption{Results from the maximum likelihood analysis for the population of \psrbh{} systems in the Galaxy.
            The corner plot shows the posterior distributions for the parameters in Eq.~\ref{eq:general_merger_rate} for the \psrbh{} systems in the Galaxy based on the event-based rate estimate (panel (a)) and population-based rate estimate (panel (b)) derived in \citet{ligo_nsbh_detections}. The contours represent 1-, 2-, and 3-$\sigma$ credible regions.}
            \label{fig:nsbh_max_lkl}
        \end{figure}
        
        The corner plot showing the posterior distribution for $N_{\rm obs, NSBH}$ and $\beta_{\rm NSBH}$ for the event- and population-based merger detection rates is shown in Fig.~\ref{fig:nsbh_max_lkl}. The maximum likelihood values corresponding to the event-based merger detection rate are $N_{\rm obs, NSBH, e} = 2.29 \pm 0.01$ and $\beta_{\rm NSBH, e} = 0.0483 \pm 0.0001$, while those corresponding to the population-based merger detection rate are $N_{\rm obs, NSBH, p} = 6.20 \pm 0.04$ and $\beta_{\rm NSBH, p} = 0.0473 \pm 0.0003$.
        
        \begin{figure}
            \centering
            \includegraphics[width = \columnwidth]{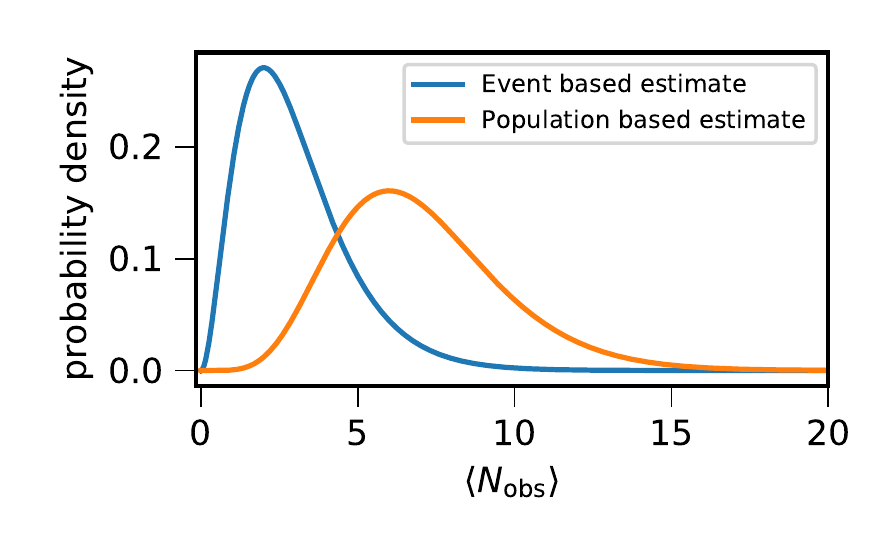}
            \caption{Distribution of the expected number of observable \psrbh{} systems from the maximum likelihood analysis described in Sec.~\ref{subsec:ml_framework}.}
            \label{fig:nsbh_nobs}
        \end{figure}
        
        These values of $N_{\rm obs, NSBH}$ can be used to estimate the average number of \psrbh{} systems, $\left< N_{\rm obs, NSBH} \right>$ that might be observable given the population parameters and radio surveys that we have considered in this work \citep[Eq.~7 in][]{kkl}. This distribution is shown in Fig.~\ref{fig:nsbh_nobs}, and predicts that we should be able to detect $\left< N_{\rm obs, NSBH, e} \right> = 2^{+5}_{-1}$ and $\left< N_{\rm obs, NSBH, p} \right> = 6^{+7}_{-4}$ if the event-based and population-based merger detection rates are accurately estimated by LIGO-Virgo respectively.
        
        The maximum likelihood value of $\beta$ which encodes the average beaming correction factor and lifetimes for the \psrbh{} population is larger than the corresponding value found for DNS systems in Sec.~\ref{subsec:test_ml_framework}. If we assume that a plausible range for beaming correction factors for canonical pulsars is $1 \leq f_b \leq 10$ \citep[][]{tm_98}, then we get a range of average lifetimes $20~{\rm Myr } \leq \tau_{\rm life} \leq 220$~Myr for the \psrbh{} systems in the Galaxy. This suggests that, similar to the DNS systems, the merger detection rate estimate might be driven by a few compact \psrbh{} binary systems.
        
    \subsection{Total number of detectable \psrbh{} systems} \label{subsec:total_psrbh_system_ligo}
        
        Similarly, we can also estimate the size of the total population of \psrbh{} that are beaming towards the Earth \citep{kkl}. For the event-based merger detection rate, the size of the population is $N_{\rm tot, e} = 77^{+202}_{-53}$, while for the population-based merger detection rate, the size of the population is $N_{\rm tot, e} = 233^{+272}_{-122}$. Given the uncertainty on the beaming correction factor from the maximum likelihood framework, it is difficult to estimate the total population of \psrbh{} systems in the Galaxy by including the systems that might not be beaming towards Earth.
        
        As discussed in Secs.~\ref{subsec:test_ml_framework} and \ref{subsec:n_psrbh_obs}, a small subset of \psrbh{} systems might be dominating the estimation of the merger detection rate. With the Galactic DNS population, there are twice as many DNS systems as ones that merge within a Hubble time, which is usually adopted as the cutoff for the lifetime of systems that are included in merger rate calculations.
        Thus, we can expect there to be more \psrbh{} systems, both observable and total, in the Galaxy than that estimated above. How many more, however, is difficult to predict since none of these systems have been found yet, and they are expected to follow a different evolutionary scenario compared to Galactic DNS systems.
        
\section{Future prospects} \label{sec:future_prospects}
    
    As we show in Sec.~\ref{sec:radio_constraint}, despite the lack of a radio detected \psrbh{} system, there could be as many as 150 \psrbh{} systems (at the 95\% credible interval) in the Galaxy that are beaming towards the Earth. On the other hand, as shown in Sec.~\ref{sec:gw_prediction}, if the LIGO-Virgo derived merger detection rate estimate is accurate, then we can expect to detect as many as 7 or 13 \psrbh{} systems (95\% credible interval), while as many as 279 or 505 (95\% credible interval) detectable \psrbh{} systems are beaming towards the Earth for the event-based and population-based rate estimates respectively.
    
    \subsection{Probability of detecting a \psrbh{} system} \label{subsec:det_prob}
        
        Given these estimates of the number of \psrbh{} systems in the Galaxy, we can calculate the probability of detecting these systems in  current radio surveys. Using the same procedure as in \citet{my_ucb_pop}, we assume that the number of \psrbh{} binaries in the Milky Way that are beaming towards Earth is given by the upper limit calculated in Sec.~\ref{subsec:stats_framework}. We choose to use this estimate so that the detection probabilities calculated here will be conservative estimates, since the detection probability is proportional to the size of the population.
        Next, we simulate $10^3$ different realizations of this population of \psrbh{} binaries using \psrpoppy, with the same prior distributions as described in Sec.~\ref{sec:radio_constraint}. We then ``run'' each of the surveys listed in Table~\ref{tab:surveys} on each of the realizations and calculate the probability with which the \psrbh{} systems are detected in any of these surveys.
        
        \begin{figure}
            \centering
            \includegraphics[width = \linewidth]{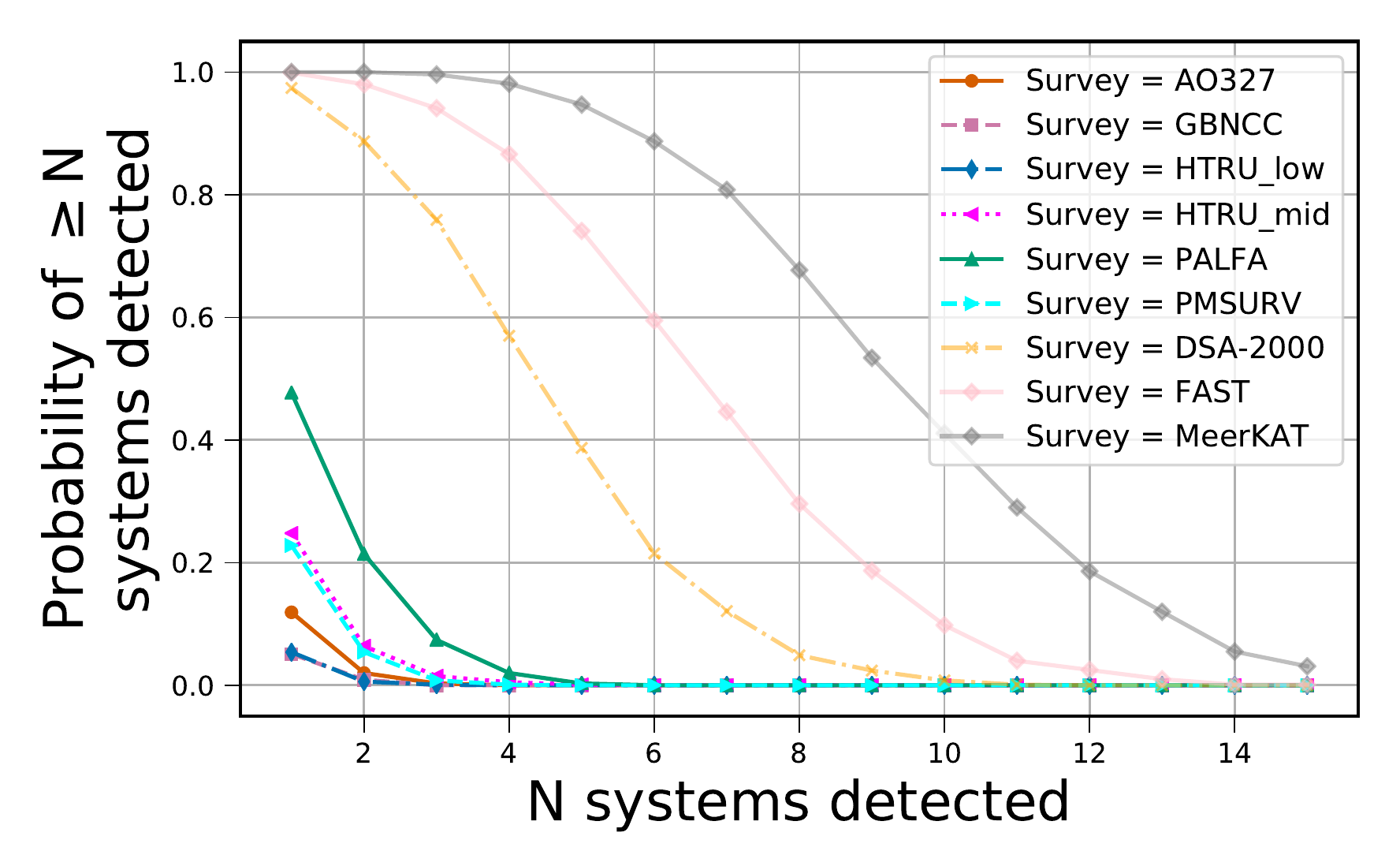}
            \caption{Probability of detecting a \psrbh{} system with  current and future radio surveys, assuming there are $\sim$150 \psrbh{} systems in the Galaxy beaming towards Earth. The PALFA survey at the Arecibo radio telescope has the highest probability of detecting a \psrbh{} system of the current surveys. Surveys underway with MeerKAT and FAST and planned with DSA-2000 should almost certainly detect at least one of these systems.}
            \label{fig:success}
        \end{figure}
        
        The probability for detection of \psrbh{} binaries for each of the current radio surveys is shown in Fig.~\ref{fig:success}. The PALFA survey has the highest probability of detecting at least one \psrbh{} system, which is to be expected given the high sensitivity of the survey. This is followed by the HTRU-mid latitude, Parkes multibeam and Arecibo driftscan surveys, while the HTRU low-latitude and GBNCC surveys have the smallest probability of detecting a \psrbh{} system. 
        
    \subsection{Where are these \psrbh{} systems?}
        
        Given the above probabilities, it is worth investigating why none of these systems have been detected in any radio pulsar surveys thus far.
        
        One reason for the lack of detection of \psrbh{} systems could be that these systems exist pre-dominantly as compact binaries, i.e. binaries with orbital periods $\sim$hours or less. There is some precedence for this assumption from ab initio stellar evolution simulations, where \psrbh{} systems in wide binaries could get disrupted in either one of the supernova explosions that form the constituent objects in the binary \citep{giacobbo_psrbh_kicks, pods_psrbh_kicks}. Detection of pulsars in binaries is hampered due to the effect of the degradation of the S/N ratio induced by the orbital motion of the pulsar in the orbit around its companion \citep[][]{og_odf, Bagchi_odf}. The orbital degradation factor \citep[][]{Bagchi_odf} is inversely proportional to the mass of the companion in the binary. Thus, pulsars in orbit around a black hole companion will suffer from larger degradation in their observed S/N relative to a pulsar in orbit around another neutron star, given the same orbital configuration \citep[see for e.g., ][]{Bagchi_odf}. However, the degradation is weaker for pulsars with larger spin periods (i.e. canonical pulsars), and we expect these pulsars to be prevalent in \psrbh{} systems. Thus, if \psrbh{} systems are in compact orbits, they could be as difficult to detect as the known Galactic DNS systems. As shown in \citet{my_ucb_pop}, the probability of detecting compact or ultra-compact pulsar binaries can be increased by using smaller survey integration times.
        
        Another hurdle towards detection of \psrbh{} systems is the relatively poor timing precision shown by canonical pulsars relative to millisecond pulsars. In particular, canonical pulsars exhibit larger red noise in their residuals along with larger measurement uncertainties \citep{young_psr_red_noise_1, young_psr_red_noise_2}, which can serve to mask the signatures of the binary motion. This phenomenon will mainly affect pulsars in wider orbits around black holes (thus only weakly influencing the merger detection rate), thereby reducing sensitivity to such \psrbh{} systems \citep{jones_in_prep}. Note that \psrbh{} systems can have, on average, orbital periods as large as $\sim$10 days and still merge within a Hubble time. Detection of pulsars in such wide \psrbh{} binaries will require a combination of multi-year monitoring of candidate canonical pulsars and higher sensitivity telescopes for conducting these timing observations \citep{jones_in_prep}.
        
        The position of the potential \psrbh{} binaries in the Galaxy could also serve to decrease the probability of detecting these systems. In particular, if \psrbh{} systems are more likely to be formed in the Galactic center \citep{psr_around_sgrA_1, psr_around_sgrA_2}, then all of the current surveys will be severely hampered by the dispersion and scattering introduced by the dense interstellar medium (ISM) present towards the Galactic center. The solution to mitigate ISM effects is to conduct surveys of the Galactic center at higher radio frequencies \citep{finding_sgrA_psr_1}, and a number of them have been carried out to date \citep{kaustubh_GC_survey, hyman_GC_survey, effelsberg_GC_survey}. However, pulsar emission tends to be weaker at higher radio frequencies \citep{bates_si_dist}, which combined with the orbital degradation effect can compound the difficulty in detecting \psrbh{} systems in the Galactic center.
        
    \subsection{New and future telescopes}
        
        The advent of a new generation of radio telescopes with larger collecting areas will allow us to probe the pulsar population at lower flux densities. Among these new instruments is the Five hundred meter Aperture Spherical Telescope \citep[FAST,][]{FAST} which is a single dish telescope offering unparalleled sensitivity, while the MeerKAT radio telescope \citep{MeerKAT}, a pre-cursor to the Square Kilometer Array \citep[SKA,][]{SKA}, and DSA-2000 \citep[Deep Synoptic Array,][]{dsa2000} are  interferometric telescopes which provide large fields of view and effective collecting areas.
        
        The pulsar survey parameters for these new instruments are listed in Table~\ref{tab:surveys} and the probability of detecting a \psrbh{} system is shown in Fig.~\ref{fig:success}.
        All of the new surveys have significantly better sensitivity than those of the current pulsar surveys, as a result of which they are almost certain to detect at least one \psrbh{} system. While the single dish configuration of FAST gives it the best sensitivity of the three new surveys, MeerKAT's access to most of the Galactic plane (given its location in the Southern hemisphere) as well as a larger field of view than FAST results in it having a higher probability of detecting more than one \psrbh{} system.
        Both of these radio telescopes have initiated surveys of the Galaxy to search for new binary pulsars \citep{FAST_gpps, meertrap} and provide the best opportunity to detect any \psrbh{} system that might have been too faint for the current radio telescopes and surveys. 
        In Fig.~\ref{fig:success}, we also illustrate detection prospects for surveys planned with the DSA-2000 concept proposed to the Astro2020 Decadal Survey \citep{dsa2000}, and show that this telescope will be competitive with FAST and MeerKAT in searching for \psrbh{} systems. Eventually the SKA will provide even better sensitivity, allowing us to probe most of the pulsar population in the Galaxy.
        
        In addition to finding new \psrbh{} systems, the higher sensitivity of these new radio telescopes also allows us to produce timing solutions with a lower root-mean-square spread ($\sigma_{\rm TOA}$) in the residuals. This improvement in the timing accuracy is important for \psrbh{} systems, since the timing accuracy determines the amount of information that can be extracted from a \psrbh{} system. For example, \citet{wex_psrbh_science} showed that for young pulsars with timing precision of $\sigma_{\rm TOA} \sim 100 \, \mu$s, any \psrbh{} system with orbital periods less than 10~days (i.e. those that are considered in this work), will allow determining the mass of the black hole to better than 5\% accuracy, while \psrbh{} systems with orbital periods less than three~days will allow measurements of the spin of the black hole within reasonable observing campaigns. As described in Sec.~\ref{subsec:total_psrbh_system_ligo}, the merger detection rate estimated using the LIGO-Virgo NS--BH mergers implies the existence of \psrbh{} systems with lifetimes that are small enough to support the existence of relatively compact \psrbh{} systems. A detection of one of these systems is thus guaranteed to provide an accurate estimate of the mass of the black hole, if not the spin of the black hole.
        
\section{Conclusion} \label{sec:conclusion}
    
    In this analysis, we use information from radio pulsar surveys and LIGO-Virgo NS--BH merger detections to estimate the number of \psrbh{} systems in the Milky Way. We find that the non-detection of a \psrbh{} system in radio pulsar surveys implies a 95\% credible upper limit of $\sim$150 \psrbh{} systems in the Galaxy that are beaming towards the Earth. Using a range distance of 100~Mpc, this corresponds to a 95\% credible upper limit on the NS--BH merger detection rate of $7.6$~yr$^{-1}$ for LIGO-Virgo. This rate is consistent with the merger detection rates derived by LIGO-Virgo \citep[][]{ligo_nsbh_detections}, implying that a non-detection of a \psrbh{} system in radio surveys so far is consistent with LIGO's detection of two NS--BH mergers.
    
    We also use the merger detection rates derived by LIGO-Virgo using their detection of two NS--BH systems to estimate the number of observable \psrbh{} systems in the Milky Way, using which we can determine the population of \psrbh{} systems that are beaming towards Earth. We find that if the LIGO-Virgo merger detection rates are accurate, there should be $\left< N_{\rm obs, NSBH, e} \right> = 2^{+5}_{-1}$ and $\left< N_{\rm obs, NSBH, p} \right> = 6^{+7}_{-4}$ detectable \psrbh{} systems in the Galaxy corresponding to the event-based and population-based merger detection rates respectively \citep[][]{ligo_nsbh_detections}. This corresponds to a total of $N_{\rm tot, e} = 77^{+202}_{-53}$ and $N_{\rm tot, e} = 233^{+272}_{-122}$ \psrbh{} systems in the Galaxy that are beaming towards the Earth.
    
    We estimate the probability of detecting these \psrbh{} systems, and show that the PALFA survey at the Arecibo radio telescope has the highest probability of detecting one of these systems. We also discuss the hurdles involved in detecting \psrbh{} systems with radio pulsar surveys and show how new and upcoming radio telescopes might help alleviate some of these hurdles, while also providing the necessary timing accuracy to enable accurate measurements of the mass and spin of the black hole in the \psrbh{} system.
    
\begin{acknowledgements}

The authors would like to thank Megan Jones for discussions on the difficulty of detecting \psrbh{} systems in wide binaries.
N.P. acknowledges support from the VIDA Fellowship from the Vanderbilt Initiative for Data-intensive Astrophysics.
N.P. and M.A.M. are members of the NANOGrav Physics
Frontiers Center (NSF PHY-1430284). M.A.M. and D.R.L. have additional support from NSF OIA-1458952. 
    
\end{acknowledgements}

\bibliography{bib.bib}
\bibliographystyle{aastex}

\end{document}